\documentclass[a4paper,11pt]{article}
\usepackage{jinstpub} 
\usepackage{chemformula}
\usepackage{graphicx}
\usepackage[separate-uncertainty]{siunitx}
\usepackage{color}
\usepackage[normalem]{ulem}
\title{\boldmath Measurement of the free ion yield and estimate of the electron mobility of Trimethyl-Bismuth (TMBi)}
\author[a]{B. Gerke}
\author[b,1]{V. Hannen\note{Corresponding author.}}
\author[b]{S. Peters}
\author[b]{C. Huhmann}
\author[c,d]{V. Sharyy}
\author[c,d]{D. Yvon}
\author[a]{K. Schäfers}
\author[b]{C. Weinheimer}
\affiliation[a]{European Institute for Molecular Imaging,
University of Münster, Röntgenstraße, Münster, Germany}
\affiliation[b]{Institute for Nuclear Physics,
University of Münster, Wilhelm-Klemm-Straße, Münster, Germany}
\affiliation[c]{IRFU, CEA, Université Paris-Saclay, Bâtiment 141, Gif-sur-Yvette, France}
\affiliation[d]{BioMAPs, Service Hospitalier Frédéric Joliot, CEA, CNRS, Inserm, Université Paris-Saclay, place du général Leclerc, Orsay, France}

\emailAdd{hannen@uni-muenster.de}

\abstract{This paper investigates the physical characteristics of the liquid Trimethyl-Bismuth (TMBi) relevant for charge readout in potential TMBi filled detectors for Positron-Emission-Tomography (PET) instruments. The high photo-efficiency and the possibility of fast timing by detection of Cherenkov light created in the liquid are both favorable for use of TMBi in a PET system. However, to obtain a high position and energy resolution also a readout of the charge signal created due to the absorption of a 511~keV photon is required. For this purpose the liquid has to be clean of electro-negative and ionic impurities. 
A distillation technique was developed within this study as one of the possible methods to remove such impurities and a measurement on the efficiency of this cleaning method is provided. The relevant quantities for an efficient charge readout are the free ion yield and the electron mobility of the liquid. We report on a new measurement of the zero-field free ion yield in TMBi and, using that result, produce an estimate of the electron mobility.
The rather low zero-field free ion yield and its much lower expected electron mobility compared to tetramethyl-n compounds like TMSi question the use of the trimethyl molecule TMBi for a PET detector liquid. But we are also giving arguments that the tetramethyl molecule TMPb could be an interesting alternative.}

\keywords{Gamma camera, PET PET/CT, Charge transport and multiplication in liquid media, Counting-gas and liquid purification}

\begin{document}
\maketitle
\flushbottom
\section{Introduction}
\begin{figure}[h]
 \centering
 \includegraphics[width=1.0\columnwidth]{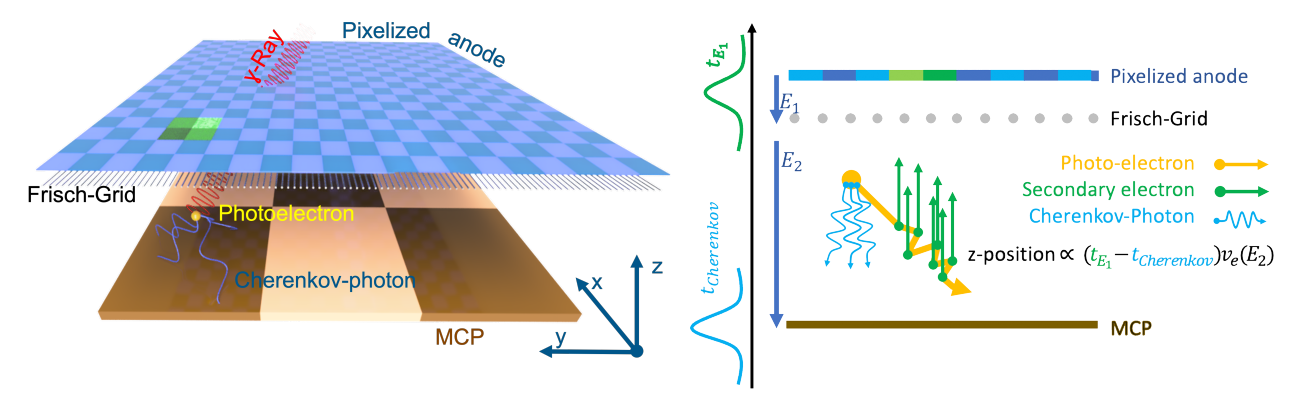}
 \caption{The proposed detector combines two readout systems to reconstruct ionization events within three dimensions and with high temporal resolution. Photo electrons created by 511 keV quanta in the medium emit Cherenkov radiation that can be detected by fast multichannel plate detectors (MCPs), providing a sharp timing signal. In addition the photo electron further ionizes the detector material, generating secondary electrons which are drifted by an applied electric field towards a segmented anode, inducing an electric signal that allows a spatial reconstruction of the event in x- and y- direction. Combining the drift time of the secondary electrons with the timing signal of the Cherenkov radiation enables determination of the z-coordinate (depth of interaction). Figure and description taken from~\cite{peters2022electro}. } 
 \label{fig:DRC}
\end{figure}

The liquid Trimethyl-Bismuth (TMBi) with a molecular mass fraction of bismuth of more than $80\,\%$ allows for a very efficient detection of 511 keV photons originating from positron annihilation, since bismuth has the highest nuclear charge (Z = 83) and thus the largest photoelectric cross section of all stable isotopes. TMBi has therefore been proposed as a promising candidate for a novel PET detector due to its high efficiency for gamma detection, its high density and the possibility for fast timing measurements using Cherenkov light produced in the liquid after absorption of a 511~keV photon~\cite{Yvon2014}. The latter aspect was studied successfully by Chyzh et al.~\cite{Chyzh_2024}. 
The additional readout of the charge signal produced by the primary photoelectron should allow for a fine grained position reconstruction. Readout of charge signals has already been realized in detectors using TMSi as target material~\cite{degrange1992,engler1999}, but not yet demonstrated with TMBi.\\
In nearly $50\,\%$ of the cases, the gamma energy is transferred to electrons in this material through the photoelectric effect. TMBi’s photo-fraction (the 511 keV photoelectric cross-section divided by the total cross-section) is the highest available, with twice the value of crystals used in modern PET systems, such as lutetium oxyorthosilicate (LSO) or lutetium yttrium oxyorthosilicate (LYSO)~\cite{Yvon2014} \cite{terWeele2015}. The general idea of detection is to detect the Cherenkov light created by the relativistic primary photoelectron using fast photo-detectors in a liquid TMBi-filled detector, supplemented by the measurement of the secondary charge carriers created by the primary photoelectron (see fig.~\ref{fig:DRC}) in ionization-chamber mode.
For reconstruction of the interaction vertex the secondary electrons are planned to be drifted towards a pixelated readout anode by applying a strong electric field, thus delivering the transversal $x$ and $y$ coordinates. The z-coordinate (along the drift direction) in the detector can then be determined from the time difference of the Cherenkov light and the drifted charge signal. \\
Together with a position sensitive readout provided by the combination of light and charge measurements, it should allow to significantly improve the contrast of the reconstructed PET image~\cite{Yvon2014}. It should be noted that a similar organo-metallic liquid, TMSi, has been used in nuclear and particle physics as a detector medium for a long time (see e.g. ref.~\cite{Engler1984}). Unfortunately it is not suited as a detector for 511~keV gammas due to the much smaller charge number ($Z=14$) of its central atom silicon. In contrast, TMBi with its high photoelectric efficiency has not been investigated as detector medium in detail so far.

In this publication we will present a new measurement of the free ion yield $G^0_{\rm fi}$ at zero field for TMBi and an estimate of the electron mobility of TMBi using the measured $G^0_{\rm fi}$ value to draw conclusions on its suitability as a PET detector. This paper is organized as follows: We start in section 2 with a discussion on the purification of TMBi and investigate distillation as a new cleaning method for TMBi. In section 3 we present our measurement of the free ion yield in TMBi (and compare it to former measurements in TMSi), from which we derive in section 4 an estimate of the electron mobility in the liquid. In section 5 we discuss the suitability of TMBi for PET applications before we conclude in section 6.
\section{Purification of TMBi by distillation}
\label{sec:distillation}
\begin{figure}[ht!]
\centering
\includegraphics[width=1 \columnwidth]{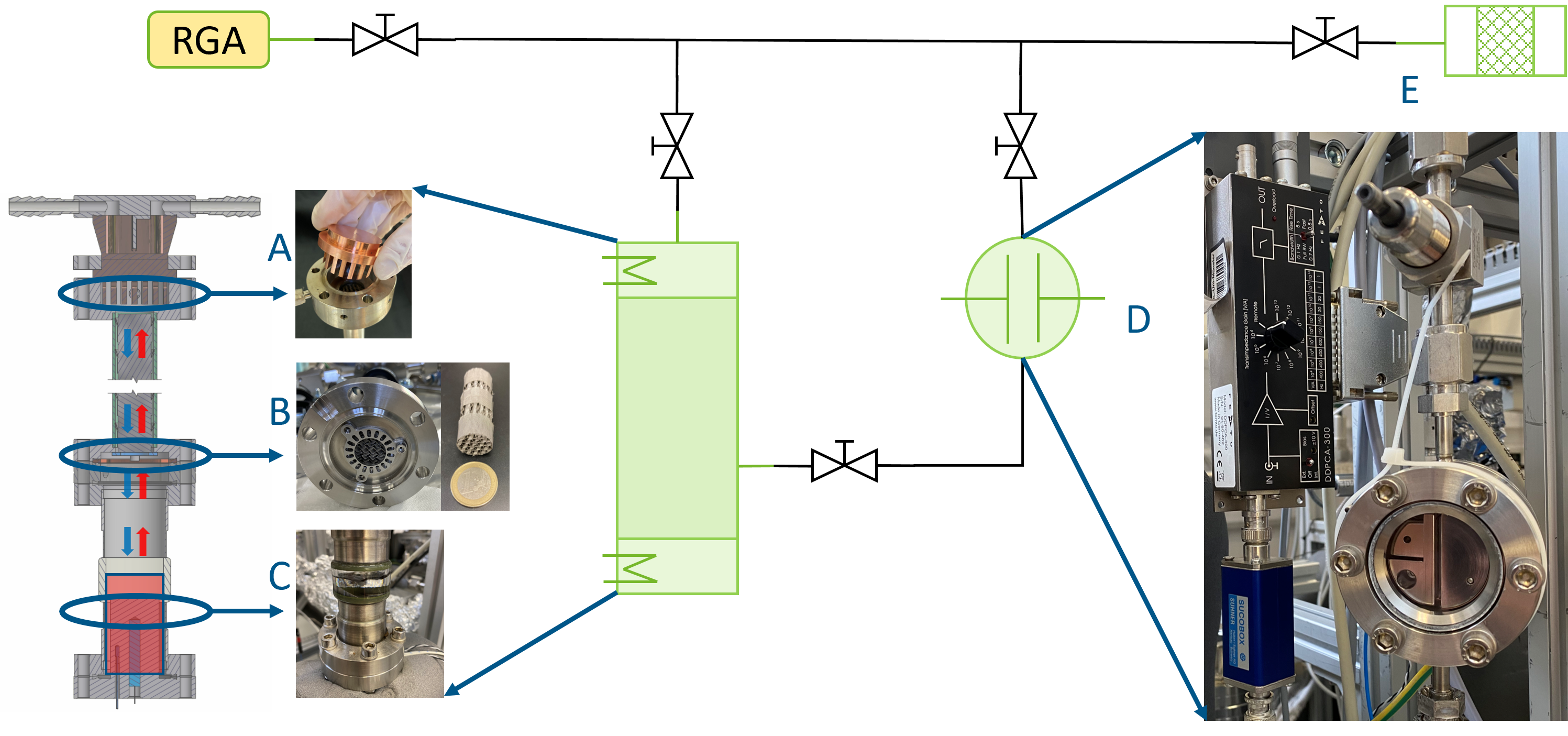}
\caption{Schematic view of the distillation purification bench. The main system consists of the rectification column (A, B, C), a test detector (D), and a residual gas analyzer (RGA). A pre-pump, turbo molecular pump, and argon filling section (E) are used to clean the system before filling with TMBi.}
\label{fig:schema}
\end{figure}

The TMBi initially used in our experiments in Münster was obtained from Chemtura (Connecticut, USA) with a claimed purity of >$99.9~\%$. However, upon delivery, the liquid appeared milky, suggesting potential contamination. To resolve this issue, we applied two purification methods: First, the TMBi was pre-filtered using molecular sieves, as previously demonstrated by the French group in our consortium and described by Farrad{\`e}che et al.~\cite{farradeche2018}. 
Second, we applied electrostatic filtering, a new idea from our XENON collaborators from Japan. We qualified the cleaning steps by measuring the dark current in a dedicated ionization chamber (see also figure~\ref{fig:detector}). These developments and investigations are described in ref.~\cite{peters2022electro}. 
While a reduction in dark current was observed with the applied cleaning steps, the amount of ionic impurities remained still too high for measurements of individual charge pulses, in contrast to our successful measurements with a TMSi filled detector cell~\cite{peters2019}. This result means that we succeeded in getting rid of some contamination but our methods did not allow to remove all relevant impurities in that batch of TMBi. For that reason we investigated distillation as a third purification method to get rid of the remaining impurities, as reported in the following.\\
The distillation purification bench (see fig.~\ref{fig:schema}) consists of a condenser (A), a packed column (B) filled with four pieces of \SI{5}{cm} Sulzer laboratory packings EX (Sulzer Management AG, Winterthur, Switzerland), a reboiler (C) heated by a \SI{100}{W} cartridge heater (WEMA, Lüdenscheid, Germany), and a pure argon filling port (E) via a getter (PS2-GC50, Pure Gase Technologies Inc., San Louis Obispo, CA, USA). The distillate can be collected in a test detector (D) or a double-walled stainless steel tank. The gas phase can be analyzed for impurities using a residual gas analyzer (RGA).\\

The setup for determining liquid purity consists of a test detector, shown on the right photograph in figure~\ref{fig:schema} and also schematically in figure~\ref{fig:detector}, a voltage source, and a Femto DDPCA-300 current amplifier read out by an NI-9205 module. The detector has a total volume of $\approx$\SI{20}{ml}. Liquid TMBi is guided directly through the \SI{2}{mm} gap between the anode and cathode (active volume $\approx$\SI{1}{ml}), enabling online conductance measurements provided the liquid is pumped through the detector.\\
For evacuation of the purification bench a combination of a pre-pump and a turbo molecular pump is used. A cold trap operated with liquid nitrogen protects the pump system from residues of TMBi. Before filling with TMBi, the system is cleaned by injecting high-purity argon and evacuating it with a turbo molecular pump to a pressure of $< \SI{1E-09}{mbar}$ and heated for more than \SI{24}{h} at temperatures up to \SI{400}{\celsius}. The system is then purged by using a hot argon stream. Infilling argon (5.0) is purified to better than \SI{10}{ppb} via a getter (E) (PS2-GC50, Pure Gase Technologies Inc., San Louis Obispo, CA, USA). After purging, the system is evacuated again to a pressure of $< \SI{1E-09}{mbar}$ for several days.\\
The purification of TMBi is performed by rectification, which relies on the establishment of theoretical distillation stages through the countercurrent interaction of rising vapor and descending liquid under equilibrium pressure conditions \cite{aprile2017removing}. The separation is driven by the differing vapor pressures of TMBi and its impurities. To enhance column efficiency, Sulzer EX laboratory packings are employed, providing multiple virtual distillation stages. Nevertheless, this level of separation does not yet yield the required TMBi purity in the parts-per-billion (ppb) range.
The efficiency of the purification procedure is monitored using a test detector, which measures the dark current with a high-sensitivity sub-femtoampere current amplifier, featuring a baseline noise of \SI{0.4}{fA} peak-to-peak (DDPCA-300, FEMTO Messtechnik GmbH, Berlin, Germany). In this context, the dark current serves as a proxy for contamination, allowing the degree of purity to be quantitatively assessed. 
Precise temperature control of the reboiler and condenser enabled the maintenance of consistent pressure conditions during distillation, which is crucial for achieving stable distillation stages and high purification efficiencies. \\
At an applied voltage of 9V at the test detector, the dark current was reduced by nearly one order of magnitude, from $ \SI{2E-09}{A}$ to a value of $ \SI{3E-10}{A}$.
A subsequent distillation step, following a comprehensive cleaning of the purification system, yielded a further decrease in dark current to $ \SI{1E-10}{A}$ yielding for both distillation steps an improvement of a factor of 20. However, the dark current in the ionization chamber and thus the conductivity of the TMBi remained elevated, with a value still a few  orders of magnitude above the requirement for operation at applied voltages of several kV. Further results on the purification analysis can be found in Appendix~\ref{app:distillation}.
\section{Determination of the free ion yield}
\label{sec:gfi}
\begin{figure}[h]
    \centering
    \includegraphics[width=0.8\textwidth]{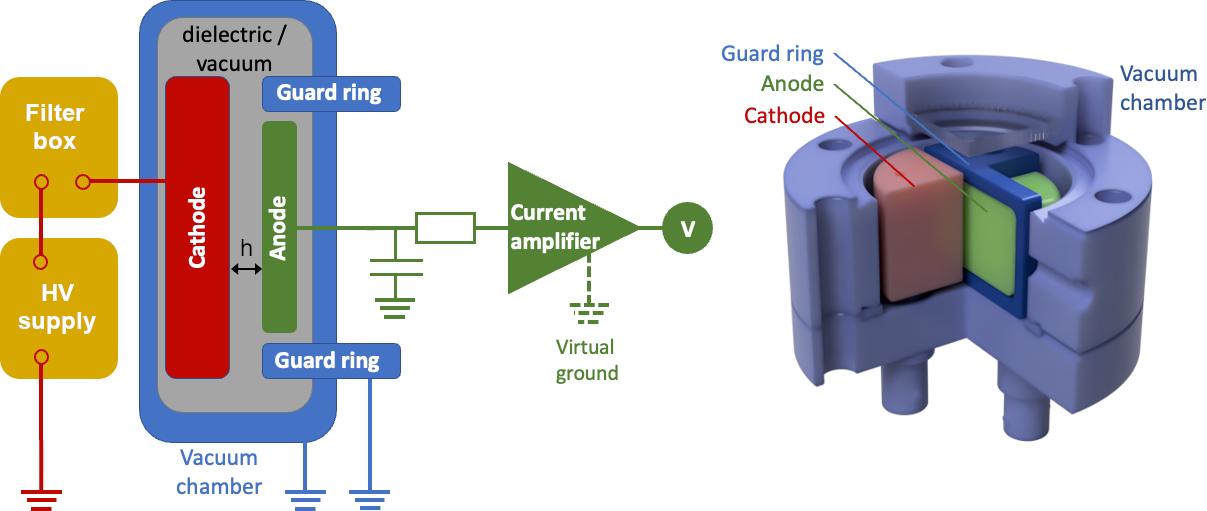}
    \caption{Detailed overview of the detector system used for $G_{\rm fi}$ and TMBi dark current measurements. Left: Schematic of the electrical readout and biasing configuration. A high-voltage (HV) supply biases the cathode via a dedicated filter box to minimize voltage ripples. The induced leakage current across the liquid gap ($h = 2\text{~mm}$) is collected at the anode, which is shielded by grounded guard rings to eliminate parasitic surface currents and guarantee a uniform electric field. The signal is processed by a high-gain current amplifier operating at virtual ground. Right: 3D CAD cross-sectional view of the physical detector cell assembly, illustrating the concentric arrangement of the cathode (red), anode (green), and insulating guard ring structure embedded within the grounded stainless-steel vacuum housing. The fluidic inlets and outlets at the bottom enable continuous online conductance monitoring of the liquid TMBi phase.}
    \label{fig:detector}
\end{figure}

\begin{figure}
 \centering
 \includegraphics[width=.8\textwidth]{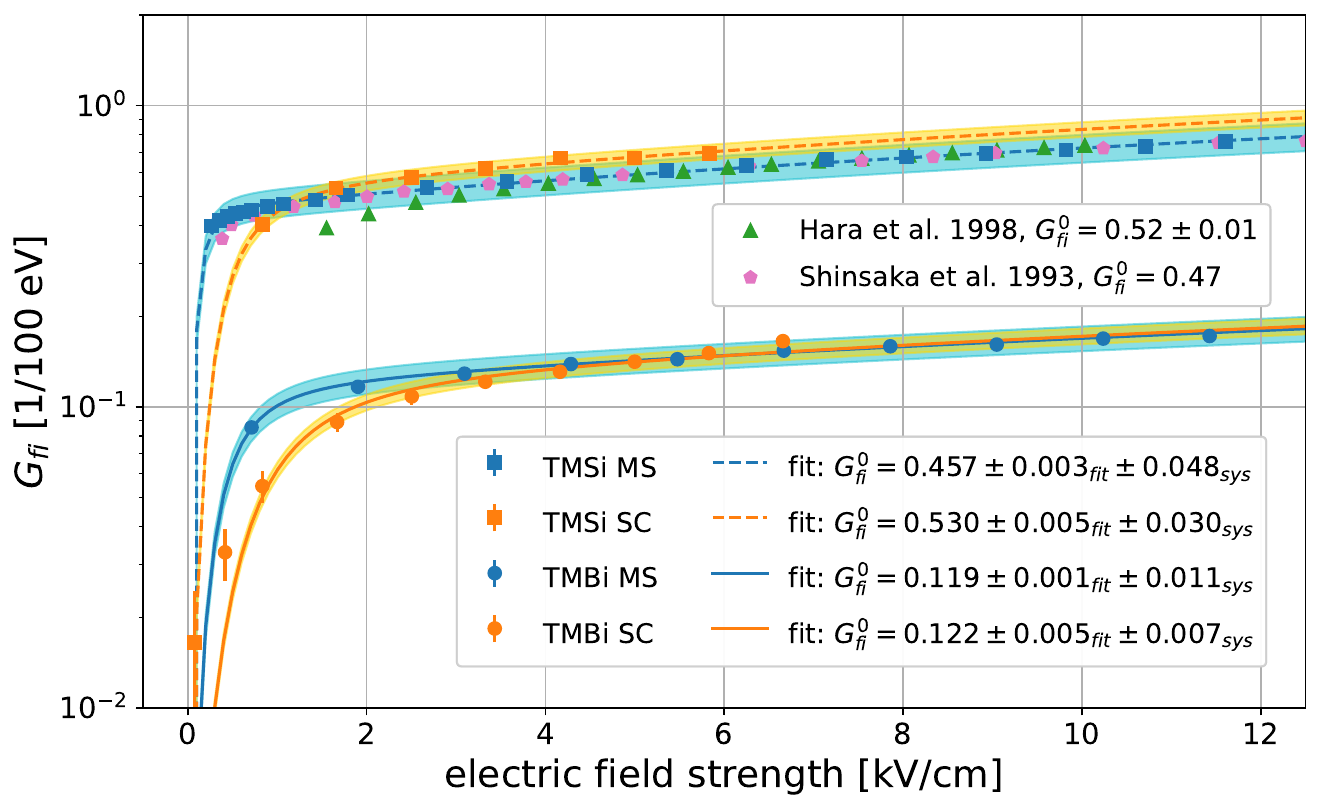}
 \caption{Free ion yield measurements performed in Münster (index MS) and Saclay~\cite{farradeche2018} (index SC) for TMBi compared to TMSi. The solid and dashed lines correspond to fit results obtained for the individual measurements. The shaded error bands are due to uncertainties of the energy dose rate used in the individual measurements. TMSi datasets from Hara et al.~\cite{Hara1998} and Shinsaka et al.~\cite{Shin1993} have been added for comparison.}
 \label{fig:FreeIonYield}
\end{figure}

In addition to being a high-Z material with a high density and corresponding high photo-fraction, there are several other physical quantities of TMBi which determine its usability as a target material for PET detectors.
Besides the relative permittivity $\epsilon_r$, the free ion yield $G_{\rm fi}$ and the electron mobility $\mu_e$ are of special importance for the detection process.\\
While we were able to measure the relative permittivity of TMBi using the material procured in Münster~\cite{peters2022electro}, the free ion yield and electron mobility could not be determined with our sample due to the high dark current observed when applying electric fields to the liquid. 
Thankfully, towards the end of our study phase, our french colleagues lend us ca. 1\;kg of ''electronic'' grade TMBi with a level of impurity less than 5~ppm~\cite{farradeche2018}, purchased from the supplier JSC~Alkyl\footnote{JSC Alkyl, 124489, Moscow, Zelenograd, proezd 4807, dom 3, stroenie 1, Russia}. This material was produced with a different method of synthesis than used by Chemtura. 
A measurement of the voltage dependent dark current resulted in a resistivity of the TMBi sample from Chemtura of $\rho \approx 1.1\cdot 10^{12}\;\Omega {\rm cm}$ while the material purchased by Saclay was found to have a resistivity of $\rho \approx 4.5\cdot 10^{14}\;\Omega {\rm cm}$.
With this cleaner TMBi sample, we were able to measure the free ion yield of TMBi over a large range of electric field strengths, which we present in this section.\\
The free ion yield $G_{\rm fi}$ is defined as the number of ion-electron pairs escaping recombination per 100~eV energy deposited in a medium. The $G_{\rm fi}$ value increases with the applied electric field $E$ and can for large field strength be approximated by~\cite{pardo2009}
\begin{equation}
 G_{\rm fi} \approx G^0_{\rm fi} \left( 1 + a E \right)  ~~ {\rm with} ~~
 a = \frac{e^3}{8\pi\epsilon_0\epsilon_r k^2_B T^2}
 \label{eq:gfi}
\end{equation}
where $G^0_{\rm fi}$ is the extrapolated free ion yield at zero electric field strength, $\epsilon_r$ the relative permittivity and $T$ the temperature of the medium. 
To take into account the so-called general recombination of charge carriers during the transport through the medium, which is responsible for an observed  drop in $G_{\rm fi}$ at low field strength, we have to multiply a correction function $f_g$ to equation~\ref{eq:gfi}. It is given by~\cite{pardo2009}
\begin{equation}
 f_g = \left[ 1 + \frac{C (1+aE)}{E^2}\right]^{-1} ~~ {\rm with} ~~ 
 C = \frac{\alpha N_{\rm ion} h^2}{6\mu_+ \mu_-}
 \label{eq:recomb_corr}
\end{equation}
where $h$ is the detector electrode distance, $\mu_+$ and $\mu_-$ are the mobilities of the positive- and negative-charge carriers, $N_{\rm ion}$ is the number of ion-electron pairs per unit volume and unit time that escaped initial recombination, and $\alpha$ is the medium-dependent recombination coefficient in units of volume per time~\cite{greening1964}.\\
To experimentally determine the free ion yield, the current $I$ induced by ionizing radiation in the detector cell is measured as a function of the applied electric field. The $G_{\rm fi}$ values are calculated using
\begin{equation}
 G_{\rm fi} = \frac{I}{e\;D}
 \label{eq:gfi_exp}
\end{equation}
with $D$ being the energy deposition rate (dose rate) in the medium in units of $(100~{\rm eV/s})$. 
The measurement setup and detector geometry used in the TMBi measurements in Münster are shown in figure~\ref{fig:detector}.
A negative high voltage is generated by an iseg~224M high voltage supply with a peak-to-peak ripple smaller than 2~mV. 
The voltage is further smoothed by a low pass filter box and then transferred to the cathode. The anode at distance $h$ from the cathode is surrounded by a grounded guard ring to reduce fringe fields at the edges of the anode. 
Signals from the anode are passed through a 1~G$\Omega$ resistor with a 220~pF capacitor to ground, protecting the measurement electronics against possible discharges. 
The current is amplified using a FEMTO DDPCA-300 current amplifier with a maximum gain of $10^{13}$V/A and a peak to peak noise of $\leq 0.4$~fA. The amplifier converts the current into an equivalent voltage which is then measured using a National Instruments NI-9205 module with a residual gain error of 115~ppm of reading. The electrode setup is housed in a CF40 enclosure, minimizing the amound of liquid required to perform the measurements.
For the TMSi measurements an older version of the detector cell was used, requiring a larger amount of liquid but using the same electrical setup.\\
Figure~\ref{fig:FreeIonYield} displays measurements of 
the free ion yields for TMSi and TMBi as detector medium performed in Saclay~\cite{farradeche2018} (index SC) and in Münster (index MS). In the Münster measurements 661~keV photons from a $^{137}$Cs source with an activity of 1.28(10)~MBq during the TMSi measurements and 1.16(9)~MBq during the TMBi measurements have been used to ionize the medium. The dose rate deposited in the active volume between the electrodes of the detector cell has been simulated using the GATE software~\cite{Jan2004}. An estimate of the simulation uncertainty has been obtained by varying the source position in the simulated geometry by $\pm 1$~mm, yielding a dose rate variation of 7\% in the TMSi and 5\% in the TMBi case.
The $G^0_{\rm fi}$ values of the individual measurements are extracted from a fit of $G_{\rm fi} \approx G^0_{\rm fi} \cdot \left( 1 + a E \right) \cdot f_g$ to the data.
In the fit constraints were put on the known relative permittivities of TMSi and TMBi of $\epsilon_{r}({\rm TMSi}) = 1.92(1)$~\cite{emob_vs_gfi} and $\epsilon_{r}({\rm TMSi}) = 2.59(3)$~\cite{peters2022electro}, respectively, and on a laboratory temperature of $T=\SI{22 \pm 2}{\celsius}$ which both enter the description via the slope $a$ in equation~\ref{eq:gfi}.
To account for some unknown fluctuations, the uncertainties of the individual data points in each fit were scaled to achieve reduced $\chi^2$ values equal to 1. The uncertainties of the $^{137}$Cs source activity and of the simulations reported above are taken into account as systematic uncertainties in the Münster fit results while the systematic uncertainties of the Saclay data were obtained from reference~\cite{farradeche2018}. Table~\ref{tab:fitresults} provides an overview of the obtained $G^0_{\rm fi}$ values.
%
%
\begin{table}[h]
 \centering
 \begin{tabular}{l|cc}
      Dataset & $G^0_{\rm fi}(\mathrm{TMSi}$) [1/100 eV] & $G^0_{\rm fi}(\mathrm{TMBi}$) [1/100 eV] \\
      \hline
    Münster & $0.457 \pm 0.003_{\rm fit} \pm 0.048_{\rm sys}$ & $0.119 \pm 0.001_{\rm fit} \pm 0.011_{\rm sys}$ \\
    Saclay  & $0.530 \pm 0.005_{\rm fit} \pm 0.030_{\rm sys}$ & $0.122 \pm 0.005_{\rm fit} \pm 0.007_{\rm sys}$\\ 
    \hline
    Error weighted mean & $0.509 \pm 0.026$ & $0.120 \pm 0.007$ \\
 \end{tabular}
 \caption{$G^0_{\rm fi}$ values for TMSi and TMBi obtained from fits to measurement data from Münster and Saclay~\cite{farradeche2018}, respectively.}
 \label{tab:fitresults}
\end{table}

While the new fit result of the $G^0_{\rm fi}$ value for TMSi data taken at Saclay reproduces the older value from reference~\cite{farradeche2018}, the result of the new fit of the Saclay TMBi free ion yield is significantly higher compared to $(0.083 \pm 0.006)\,/\,100\,{\rm eV}$ from ref.~\cite{farradeche2018}. This discrepancy can be traced back to the fact that in the old fit neither the relative permittivity of TMBi, which was not known at that time, nor the temperature of the measurement setup were constrained. 
While there are some differences in the TMSi results obtained in Münster and Saclay, caused by the different experimental setups, impurity concentrations in the media and deviations in the simulation of the dose rates, the fit results for the free ion yields are consistent within the uncertainties.
In particular, the far-reaching conclusion that $G^0_{\rm fi}({\rm TMBi})$ is more than a factor 4 smaller than the $G^0_{\rm fi}({\rm TMSi})$ is confirmed by the recent Münster data.
The difference in the free ion yield between TMSi and TMBi will lead to significantly smaller charge signals in a TMBi filled detector, challenging its application as a gamma detecting medium.
\section{Estimation of the electron mobility}
\begin{figure}[h]
 \centering
 \includegraphics[width=.8\textwidth]{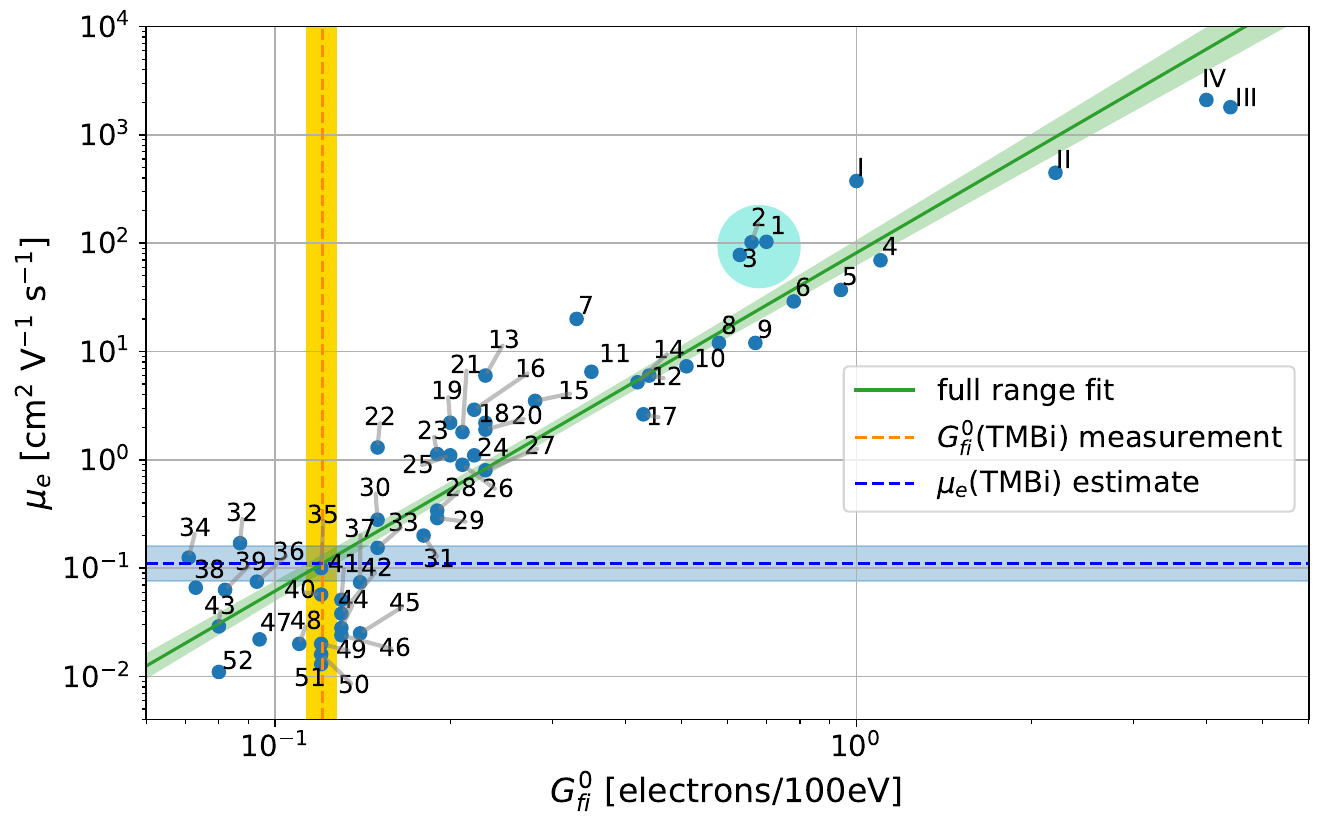}
 \caption{\label{fig:emob_vs_gfi} Electron mobility $\mu_e$ vs. zero-field free ion yield $G^0_{\rm fi}$ of dielectric liquids from ref.~\cite{emob_vs_gfi}. Room temperature liquids are labeled with numbers, cryogenic liquids with Roman numerals. The cyan circle highlights methyl compounds from the 4th main group of the periodic table (same as TMSi and TMPb). The green line and uncertainty band correspond to the fit result of eq.~\ref{eq:prop_mu_gfi}. The yellow band indicates the uncertainty interval of the measured $G^0_{\rm fi}$ value for TMBi. The blue band shows the resulting uncertainty interval for the estimate of the TMBi electron mobility $\mu_e$. The compound names are listed below:\\
 \textbf{1}: Tetramethylsilane,  \textbf{2}: Tetramethylgermane, 
 \textbf{3}: Tetramethylstannane, 
 \textbf{4}: 2,2-Dimethylpropane, \textbf{5}: Ethane, \textbf{6}: 2,2,4,4-Tetramethylpentane, \textbf{7}: Hexamethyldisilane, \textbf{8}: 2,2-Dimethylbutane, \textbf{9}: 2,2,5,5-Tetramethylhexane, \textbf{10}: 2-Methylpropane, \textbf{11}: 2,2,4-Trimethylpentane (isooctane), \textbf{12}: 2,3-Dimethylbutene-2, \textbf{13}: 1,4-Cyclohexadiene, \textbf{14}: 2,2,3,3-Tetramethylpentane, \textbf{15}: 4,4-Dimethyl-cis-2-pentene, \textbf{16}: 2-Methyl-2-butene, \textbf{17}: Propane, \textbf{18}: cis-Butene-2, \textbf{19}: Tetrakis ethylene, \textbf{20}: 3,3-Dimethyl-1-butene, \textbf{21}: 4,4-Dimethyl-trans-2-pentene, \textbf{22}: Cyclopentene, \textbf{23}: Cyclopentane, \textbf{24}: 2,3-Dimethylbutane, \textbf{25}: Cyclohexene, \textbf{26}: 2-Methylbutane, \textbf{27}: 3,3-Diethylpentane, \textbf{28}: n-Butane, \textbf{29}: 2-Methylpentane, \textbf{30}: Cyclohexane, \textbf{31}: 3-Methylpentane, \textbf{32}: 1,3,5-Trimethylbenzene, \textbf{33}: n-Pentane, \textbf{34}: Benzene, \textbf{35}: cis-Decalin, \textbf{36}: Methylbenzene, \textbf{37}: n-Hexane, \textbf{38}: 1,4-Dimethylbenzene, \textbf{39}: 1.3-Dimethylbenzene, \textbf{40}: Methylcyclohexane, \textbf{41}: n-Heptane, \textbf{42}: n-Octane, \textbf{43}: trans-Butene-2, \textbf{44}: n-Nonane, \textbf{45}: n-Decane, \textbf{46}: n-Undecane, \textbf{47}: 1,2-Dimethylbenzene, \textbf{48}: 1,2,3,4-Tetramethylbenzend, \textbf{49}: n-Dodecane, \textbf{50}: n-Tetradecane, \textbf{51}: trans-Decalin, \textbf{52}: Hexafluorobenzene, 
 \textbf{I}: Methane(\SI{122}{K}), \textbf{II}: Argon(\SI{95}{K}), \textbf{III}: Xenon(\SI{164}{K}), \textbf{IV}: Krypton (\SI{129}{K}).} 
\end{figure}

While the free ion yield determines the number of charge carriers created per deposited energy and therefore the magnitude of the measurable signal, the electron mobility determines not only the time resolution, and via the trigger by the Cherenkov signal the event position in the detector, but it also governs the dead time of a TMBi based detector and therefore limits its rate capability.
Though the electron mobility in TMBi has not yet been measured, an estimate can be obtained from the relation observed between known $(G^0_{\rm fi}\; , \;\mu_e)$ pairs from other dielectric liquids. A large number of experimental values have been collected by Jay-Gerin et al.~\cite{emob_vs_gfi} who found that on a double logarithmic scale, as displayed in figure~\ref{fig:emob_vs_gfi}, the data points approximately follow a linear function.
From this Jay-Gerin et al. deduced, that the relation between the free ion yield at zero electric field and the electron mobility can be described by
\begin{align}
  G^0_{\rm fi} \approx \beta (\mu_{e})^n 
  ~\Leftrightarrow~ 
  \mu_e \approx (G^0_{\rm fi} / \beta)^{1/n} ~ .
  \label{eq:prop_mu_gfi}
\end{align} 
A fit of this equation to a subset of the data with electron mobilities $\mu_{e}>\SI{0.1}{cm^2 V^{-1}s^{-1}}$ performed in ref.~\cite{emob_vs_gfi} delivered $\beta = 0.21(2)$ and $n = 0.31(5)$.
As from the measured $G^0_{\rm fi}$ value we expect an electron mobility for TMBi that may even be below $\SI{0.1}{cm^2 V^{-1}s^{-1}}$, we performed an updated fit of equation~\ref{eq:prop_mu_gfi} to all data points, resulting in $\beta = 0.244(21)$ and $n = 0.320(16)$ shown as green line and uncertainty band in figure~\ref{fig:emob_vs_gfi}.
Using these numbers we can estimate a range of possible electron mobilities for TMBi by evaluating equation~\ref{eq:prop_mu_gfi} for the mean $G^0_{\rm fi}({\rm TMBi})$ value determined from the Münster and Saclay datasets (see table~\ref{tab:fitresults}) shown as yellow band in the figure. We obtain
\begin{equation}
 \mu_{e}({\rm TMBi}) = 0.110^{+0.049}_{-0.034} \; {\rm \frac{cm^2}{V s}}\; .
\end{equation}
The estimated electron mobility in TMBi is about three orders of magnitude lower than in Tetramethyl-n compounds like TMSi. This rather low mobility value will lead a low time resolution and to dead times of TMBi based detectors which are consequently three orders of magnitude longer than for a TMSi filled detector.
\section{Conclusion}
The tests and measurements performed in this work aim at a purification of TMBi for use in a possible PET detector and a characterization of the relevant quantities for charge detection. For use as a detector medium with charge readout, additional physical characteristics of the target material become important besides its high photo-efficiency, namely stability against discharges, the free ion yield and the electron mobility in the liquid.\\
The problem of rapid decomposition of TMBi under certain circumstances that was encountered in Saclay, was solved by pressurizing the detectors with a purified argon atmosphere. This allows safe handling of the liquid when applying large electric fields (see ref.~\cite{gerke2022suppression}).\\
A measurement of the free ion yield was initially tried with TMBi procured from the company Chemtura. Several cleaning methods were tested on the liquid to reduce the large dark current observed when applying electric fields to the material: molecular sieves (which were used successfully in Saclay), distillation (inspired by our successful distillation systems for the XENON dark matter experiments) described in section~\ref{sec:distillation} and electrostatic filtering, as described in ref.~\cite{peters2022electro}.
The cleaning by distillation, reported on in this work, lead to a reduction in the dark current measured in the TMBi filled detector cell by a factor 20.
While these methods enabled a measurement of the relative permittivity $\epsilon_r$ of the material~\cite{peters2022electro}, the amount of electronegative or ionic impurities remained too high to determine the free ion yield or detect individual charge pulses like we were able to do with a TMSi filled detector cell~\cite{peters2019}.\\
A TMBi sample produced by a different method and provided to us by our french colleagues proved to be much cleaner. A measurement of the voltage dependent dark current resulted in a resistivity of the TMBi sample from Saclay which was a factor $\approx 400$ higher than the material purchased from Chemtura.
With the cleaner sample, we were able to determine the zero-field free ion yield $G^0_{\rm fi}$ of TMBi from free ion yield $G_{\rm fi}$ measurements over a large range of electric field strengths and obtain a good agreement with an updated fit of measurements performed in Saclay. 
The datasets from Münster and Saclay give a combined value of $G^0_{\rm fi}(\mathrm{TMBi}) = (0.120 \pm 0.007)\,/\,100\;{\rm eV}$ compared to a value of $G^0_{\rm fi}(\mathrm{TMSi}) = (0.509 \pm 0.026)\,/\,100\;{\rm eV}$ for TMSi.
The zero-field free ion yield of TMBi is therefore about a factor 4.2 lower than that of TMSi, reducing the amplitude of the observable charge signals in TMBi filled detectors accordingly.\\
Using the measured $G^0_{\rm fi}$ value it is possible to estimate the electron mobility $\mu_e$ in TMBi from the systematics of published electron mobilities for other dielectric liquids~\cite{emob_vs_gfi}. We obtain a value of $\mu_{e}({\rm TMBi}) = 0.110^{+0.049}_{-0.034} \; {\rm \frac{cm^2}{V s}}$ that is about three orders of magnitude smaller than the electron mobility in Tetramethyl-n compounds like TMSi. This rather low mobility value will lead to a low time resolution and to dead times of TMBi based detectors which are consequently three orders of magnitude longer than for a TMSi filled detector.\\
Both the low zero-field free ion yield, leading to small charge signals, and the low electron mobility in the liquid, leading to a low time resolution and large dead times, disfavor the use of TMBi as a medium for future PET detectors.
On the other hand Tetramethyl-n compounds, with elements from the 4$^{\rm th}$ main group of the periodic table like TMSi, TMGe or TMSn, have both a high zero-field free ion yield and a high electron mobility (visible in figure~\ref{fig:emob_vs_gfi}).
It is therefore reasonable to assume that TMPb (Tetramethyl-Lead), which features the same 4 methyl groups in a Tetrahedral configuration around the central metallic element as TMSi, to be a more promising candidate for use in a metallo-organic liquid filled PET detector, with an only slightly lower photoelectric efficiency and high free electron yield and electron mobility.
This expectation is supported by quantum chemistry calculations of the orbital configurations of the tetrametyl molecules TMSi, TMBi and TMPb~\cite{farradeche2018} that predict a global behaviour of the molecules TMSi and TMPb to donate electrons and for the trimethyl molecule TMBi to capture electrons, which is detrimental to the electron transport in the latter liquid. While the physical properties of TMPb are expected to be favorable, an experimental characterization of the liquid would require additional safeguards due to the high toxicity of the material.
\appendix
\section{Appendix: Purification analysis of distillation}
\label{app:distillation}
\begin{figure}[ht!]
 \centering
 \includegraphics[width=1.0\columnwidth]{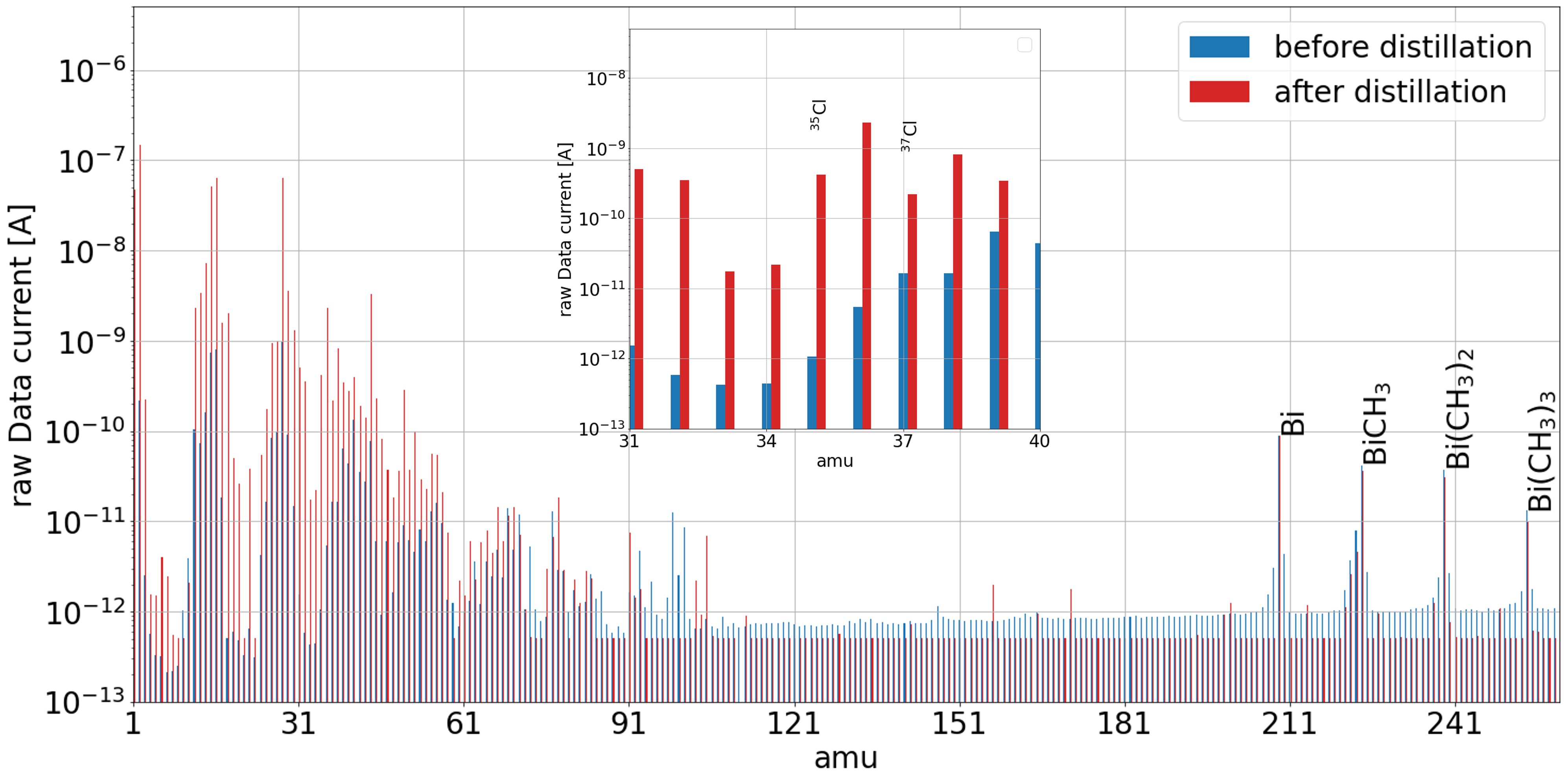}
 \caption{RGA mass spectra comparing the unpurified TMBi medium (blue) with the residual process gas remaining after distillation (red). The post-distillation data are normalized to the elemental $\text{Bi}^+$ peak ($209\text{~amu}$) of the pre-distillation spectrum to characterize mass ratio variations as a function of the bismuth content.
For both spectra, signals above 100 amu approach similar flat, mass-independent baselines, corresponding to the instrumental noise floor rather than a genuine chemical background. The high-resolution inset ($31\text{--}40\text{~amu}$) highlights a massive relative accumulation of contaminants; the signals at $36\text{~amu}$ and $38\text{~amu}$ closely track the natural $3:1$ chlorine isotope distribution, pointing toward $\text{HCl}$ as a primary constituent while accounting for potential isobaric overlaps from hydrocarbon fragments ($\text{C}_3\text{H}_x^+$) or trace argon isotopes ($^{36}\text{Ar}/^{38}\text{Ar}$) in this mass range.}
 \label{fig:RGA}
\end{figure}

To evaluate the efficiency of the distillation process for the purification of TMBi, residual gas analysis (RGA) was performed at room temperature  conditions (fig. \ref{fig:RGA}). 
The analysis compares the unpurified precursor phase (blue spectrum) with the residual process gas remaining in the vacuum system after the distilled TMBi batch had been isolated (red spectrum), i.e., comparing the non-purified TMBi not against the purified product itself, but against the contaminant-enriched left over during the separation process. To track the chemical evolution and directly characterize the variation of mass ratios independent of absolute pressure fluctuations, the post-distillation dataset was normalized by anchoring its elemental bismuth ion peak ($\text{Bi}^+$, $209\text{~amu}$) to the corresponding $\text{Bi}^+$ peak intensity of the pre-distillation reference spectrum.\\
This relative normalization directly highlights the shift in chemical composition per unit of bismuth framework. Prior to distillation, the unpurified medium exhibits the characteristic fragmentation pattern of TMBi ($>200\text{~amu}$) alongside a continuous background of intermediate synthesis byproducts distributed between $60\text{~amu}$ and $100\text{~amu}$. Above this threshold, the pre-distillation spectrum itself approaches a flat, mass-independent baseline on the order of $10^{-12}\text{--}10^{-13}\text{~A}$, which we attribute to the electronic noise floor of the RGA detector rather than to a genuine chemical background (see also fig.~\ref{fig:RGA}).\\
Following the process, the normalized residual gas spectrum (red curve) demonstrates a relative increase of contaminants 
in the range between $10$ and $70\text{~amu}$, where the pre-distillation spectrum shows peaks clearly resolvable above the instrumental noise floor.
The high-resolution inset (fig. \ref{fig:RGA}) of the $31\text{--}40\text{~amu}$ mass range strongly suggests that a major component of these residues consists of hydrogen chloride ($\text{HCl}$), although contributions from overlapping mass signals of other impurities must be considered. Specifically, the prominent features at $36\text{~amu}$ and $38\text{~amu}$ closely resemble the characteristic $3:1$ natural isotopic abundance ratio of chlorine ($^{35}\text{Cl}$ and $^{37}\text{Cl}$), indicating a significant presence of $\text{H}^{35}\text{Cl}^+$ and $\text{H}^{37}\text{Cl}^+$ ions, likely stemming from the manufacturing process as indicated in~\cite{lanxess2020TMBi}. However, to account for potential isobaric interferences in this mass region, contributions from light hydrocarbon fragments derived from methyl-group decomposition (such as $\text{C}_3\text{H}_x^+$ species like $\text{C}_3\text{H}_2^+$ at $38\text{~amu}$) or trace atmospheric noble gases (including $^{36}\text{Ar}^+$ and $^{38}\text{Ar}^+$) cannot be entirely ruled out and likely superimpose on the primary chlorine signals.\\
The primary components at mass numbers above 200 were identified as Bi (\SI{209}{amu}), \ch{Bi(CH3)} (\SI{224}{amu}), \ch{Bi(CH3)2} (\SI{239}{amu}), and \ch{Bi(CH3)3} (\SI{254}{amu}). Various hydrocarbon compounds such as ethene \ch{C2H4} (\SI{28}{amu}), ethane \ch{C2H6} (\SI{30}{amu}), and propane \ch{C3H8} (\SI{44}{amu}) appear at lower mass numbers. \\
To evaluate the purification efficiency, two samples were compared: the untreated TMBi prior to distillation and the residues remaining in the distillation column. At the same applied voltage, the distilled TMBi exhibited a significantly lower dark current than the starting material, indicating successful purification.\\
This measurement suggests that the distillation process is effective, but may not be sufficient to remove all contaminants, indicating that additional purification steps may be required to produce high-purity material. The implementation of a multi-stage distillation unit, combined with the ability to drain the reboiler’s sump, might enable further optimization of the purification process.
\acknowledgments
We would like to thank Deutsche Forschungsgemeinschaft (DFG), project numbers WE 1843/8-1 and SCHA 1447/3-1, French national research agency (ANR), project number ANR-18-CE92-0012-01.
%
\bibliographystyle{JHEP}
\bibliography{biblio.bib}
\end{document}